\documentclass[aps,prl,preprint,groupedaddress]{revtex4}

\usepackage{graphicx}
\begin{document}


\title{Efficiency optimization in a correlation ratchet with asymmetric unbiased fluctuations}


\author{Bao-Quan  Ai$^{1}$}\author{Xian-Ju Wang$^{1}$}\author{Guo-Tao  Liu$^{1}$}
 \author{De-Hua Wen$^{1, 2}$}\author{Hui-Zhang Xie$^{2}$}\author{Wei Chen$^{3}$}\author{Liang-Gang Liu$^{1}$}
\affiliation{$^{1}$ Department of Physics, ZhongShan University,
GuangZhou, China\\ $^{2}$ Department of Physics, South China
University of technology, GuangZhou, China \\
$^{3}$ Department of Physics, JiNan University, GuangZhou, China}


\date{\today}

\begin{abstract}
The efficiency of a Brownian particle moving in periodic potential
in the presence of asymmetric unbiased fluctuations is
investigated. We found that there is a regime where the efficiency
can be a peaked function of temperature,  which  proves that
thermal fluctuations facilitate the efficiency of energy
transformation, contradicting the earlier findings (H. kamegawa et
al. Phys. Rev. Lett. 80 (1998) 5251). It is also found that the
mutual interplay between asymmetry of fluctuation and  asymmetry
of the potential may induce optimized efficiency at finite
temperature. The ratchet is not most efficiency when it gives
maximum current.
\end{abstract}

\pacs{05. 40. -a, 02. 50.Ey, 87. 10. +e}

\maketitle

\section{1. Introduction}
 \indent Much of the interest in non-equilibrium induced
transport processes in concentrated on stochastically driven
rachet \cite{1}\cite{2}\cite{3}. This subject was motivated by the
challenge to explain unidirectional transport in biological
systems, as well as their potential novel technological
applications ranging from classical non-equilibrium models
\cite{4}\cite{5} to quantum systems \cite{6}\cite{7}\cite{8}.
Several models have been proposed to describe muscle's
contraction\cite{9}\cite{10}\cite{11}, or the asymmetric
polymerization of actin filaments responsible of cell
motility\cite{1}\cite{12}.

\indent Rectification of noise leading to unidirectional motion in
ratchet systems has been an active field of research over the last
decade. In these systems directed Brownian motion of particles is
induced by nonequilibrium noise in the absence of any net
macroscopic forces and potential gradients. Several physical
models have been proposed: rocking ratchets \cite{13}, fashing
ratchets \cite{14}, diffusion ratchets \cite{15}, correlation
ratchets \cite{16}, etc. In all these studies the potential is
taken to be asymmetric in space. It has also been shown that one
can obtain unidirectional current in the presence of spatially
asymmetric potentials. For these nonequilibrium systems external
random force should be time asymmetric or the presence of space
dependent mobility is required.

\indent The energetics of these systems, which rectify the
zero-mean fluctuations, are investigated in recent years
\cite{17}\cite{18}\cite{19}. To define optimal models for such
ratchet systems, the maximization of the efficiency of energy
transformation is inevitable. Much of interest  was motivated by
the elegant piece of work done by Magnasco \cite{13}, which showed
that a Brownian particle, subject to external  fluctuations, can
undergo a non-zero drift while moving under the influence of an
asymmetric potential. The temperature dependence of  the current
has been studied and it has been shown that  the current can be a
peaked function of temperature. He claimed that there is a region
where the efficiency can be optimized at finite temperatures and
the existence of thermal fluctuations facilitate the efficiency of
energy transformation. Based on energetic analysis of the same
model  Kamegawa et al. \cite{19} made a important conclusion that
the efficiency of energy transformation cannot be optimized at
finite temperatures and that the thermal fluctuations does not
facilitate it. Recently, investigation of Dan et al. \cite{17}
showed that the efficiency can be optimized at finite temperatures
in inhomogeneous systems with spatially varying friction
coefficient in an adiabatically rocked rachet, and efficiency
optimization in homogeneous nonadiabatical rachet systems was
observed by Sumithra et al.\cite{18}.  The equation of whether the
thermal fluctuations actually facilitate the energy transformation
in forced homogeneous adiabatical ratchet systems is still unknown
and it is the subject of the current investigation.
\section{2. The model}
\indent We consider a rocking ratchet system subject to an
external load:
\begin{equation}\label{1}
\frac{dx}{dt}=-\frac{\partial V_{0}(x)}{\partial x}-\frac{\partial
V_{L}(x)}{\partial x}+F(t)+\sqrt{2k_{B}T}\xi(t),
\end{equation}

\begin{figure}[htbp]
  (a)\begin{center}\includegraphics[width=11cm,height=8cm]{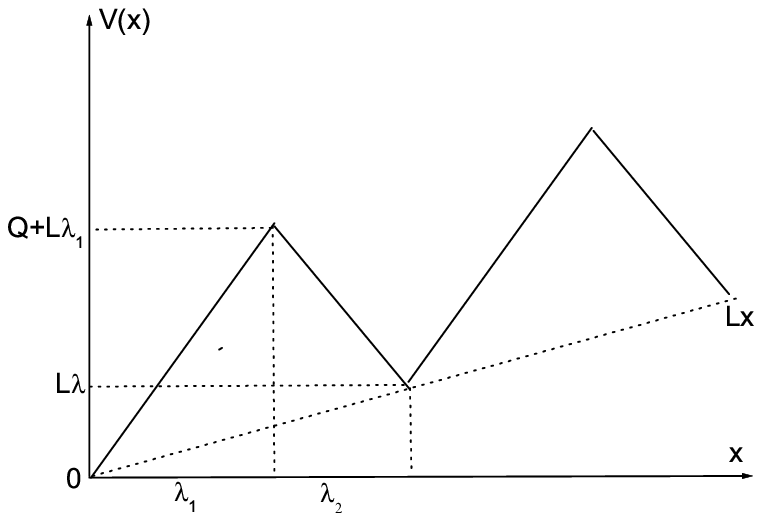}
  \end{center}
(b)
  \begin{center}
\includegraphics[width=11cm,height=8cm]{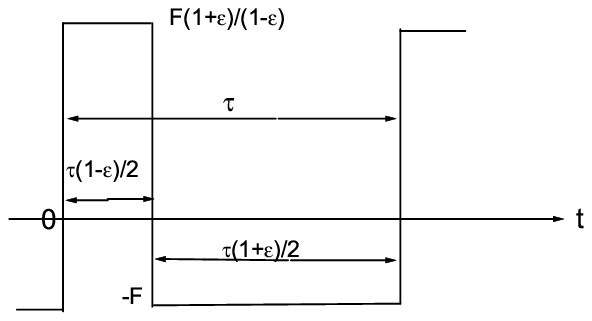}
\end{center}
  \caption{(a) Schematic illustration of the potential, $V(x)=V_{0}(x)+V_{L}(x)$, $V_{0}(x)$ is a piecewise
  linear and periodic potential. $V_{L}$ is a potential due to the load.
The period of the potential is $\lambda=\lambda_{1}+\lambda_{2}$,
and $\Delta=\lambda_{1}-\lambda_{2}$.  (b) The driving force
$F(t)$ which preserved the zero mean $<F(t)>=0$, where the
temporal asymmetry is given by the parameter $\varepsilon $.}
   \label{3}

\end{figure}

where $x$ represents the state of the ratchet. $V_{0}$ is a
periodic potential, $\xi(t)$ is a randomly fluctuating Gaussian
white noise with zero mean and with autocorrelation
function$<\xi(t)\xi(s)>=\delta(t-s)$. Here $<...>$ denotes an
ensemble average over the distribution of the fluctuating forces
$\xi(t)$.  $V_{L}$ is a potential against which the work is done
and $\frac{\partial V_{L}}{\partial x}=L>0$ is the load force. The
geometry of potential $V(x)=V_{0}(x)+V_{L}(x)$ is displayed in
Fig. 1a.  $F(t)$ is some external driving force which is shown in
Fig.1b.  The evolution of the probability density for $x$ is given
by the associated Fokker-Planck equation,
\begin{equation}\label{2}
\frac{\partial P(x,t)}{\partial t}=\frac{\partial}{\partial
    x}[k_{B}T\frac{\partial P(x,t)}{\partial
    x}+(V^{'}(x)-F(t))P(x,t)]=-\frac{\partial j}{\partial x} .
\end{equation}
If $F(t)$ changes very slowly, there exists a quasi-stationary
state. In this case, the average current of the particle can be
solved by evaluating the constants of integration under the
normalization condition and the periodicity condition of $P(x)$,
and the current can be obtained and expressed as \cite{13}
\begin{equation}\label{3}
j(F(t))=\frac{P_{2}^{2}\sinh[\lambda(F(t)-L)/2k_{B}T]}{k_{B}T(\lambda/Q)^{2}P_{3}-(\lambda/Q)P_{1}P_{2}\sinh[\lambda(F(t)-L)/2k_{B}T]},
\end{equation}
where
\begin{equation}\label{e1}
P_{1}=\Delta+\frac{\lambda^{2}-\Delta^{2}}{4}\frac{F(t)-L}{Q},
\end{equation}
\begin{equation}\label{e2}
P_{2}=(1-\frac{\Delta(F(t)-L)}{2Q})^{2}-(\frac{\lambda(F(t)-L)}{2Q})^{2},
\end{equation}
\begin{equation}\label{e3}
P_{3}=\cosh\{[Q-0.5\Delta(F(t)-L)]/k_{B}T\}-\cosh[\lambda(F(t)-L)/2k_{B}T],
\end{equation}

where $\lambda=\lambda_{1}+\lambda_{2}$ and
$\Delta=\lambda_{1}-\lambda_{2}$. The average current, the
quantity of primary interest, is given by
\begin{equation}\label{4}
J=\frac{1}{\tau}\int_{0}^{\tau}j(F(t))dt,
\end{equation}
where $\tau$ is the period of the driving force $F(t)$, which is
assumed longer than any other time scale of the system in this
adiabatic limit. Magnasco considered this case, but only for
$F(t)$ symmetric in time. Here will again consider a driving with
a zero mean, $<F(t)>=0$, but which is asymmetric in time \cite{20}
\begin{eqnarray}\label{7}
  F(t)=\frac{1+\varepsilon}{1-\varepsilon}F,               (n\tau\leq
t<n\tau+\frac{1}{2}\tau(1-\varepsilon)), \\
      =-F,    (n\tau+\frac{1}{2}\tau(1-\varepsilon)<t\leq
      (n+1)\tau).
\end{eqnarray}
\indent In this case the time averaged current is easily
calculated,
\begin{equation}\label{8}
    J=\frac{1}{2}(j_{1}+j_{2}),
\end{equation}
where
$j_{1}=(1-\varepsilon)j(\frac{1+\varepsilon}{1-\varepsilon}F)$ and
$j_{2}=(1+\varepsilon)j(-F)$.

 \indent The input energy $R$ per unit time from external
force to the ratchet and the work $W$ per unit time that the
ratchet system extracts from the fluctuation into the work are
given respectively\cite{19}:
\begin{equation}\label{9}
R=\frac{1}{t_{j}-t_{i}}\int^{x(t_{j})}_{x(t_{i})}F(t)dx(t),
\end{equation}
\begin{equation}\label{10}
W=\frac{1}{t_{j}-t_{i}}\int^{x(t_{j})}_{x(t_{i})}dV(x(t)).
\end{equation}
For the square wave, they yield:
\begin{equation}\label{11}
  <R>=\frac{1}{2}F(j_{1}-j_{2}),
\end{equation}
\begin{equation}\label{12}
<W>=\frac{1}{2}L(j_{1}+j_{2}).
\end{equation}
\indent Thus the efficiency $\eta$ of the system to transform the
external fluctuation to useful work is given by
\begin{equation}\label{13}
  \eta=\frac{<W>}{<R>}=\frac{L(j_{1}+j_{2})}{F(j_{1}-j_{2})}.
\end{equation}
which in turn, being $\frac{j_{2}}{j_{1}}<0$, can be written as
\begin{equation}\label{14}
\eta=\frac{L}{F}(\frac{1-|j_{2}/j_{1}|}{1+|j_{2}/j_{1}|}).
\end{equation}
\section{3. Results and discussion}
\indent we have calculated the efficiency and the net current as a
function of temperature $T$ for the case where asymmetric unbiased
fluctuations are applied, and the results are shown in Fig. 2-
Fig. 6.

\begin{figure}[htbp]
  \begin{center}\includegraphics[width=11cm,height=8cm]{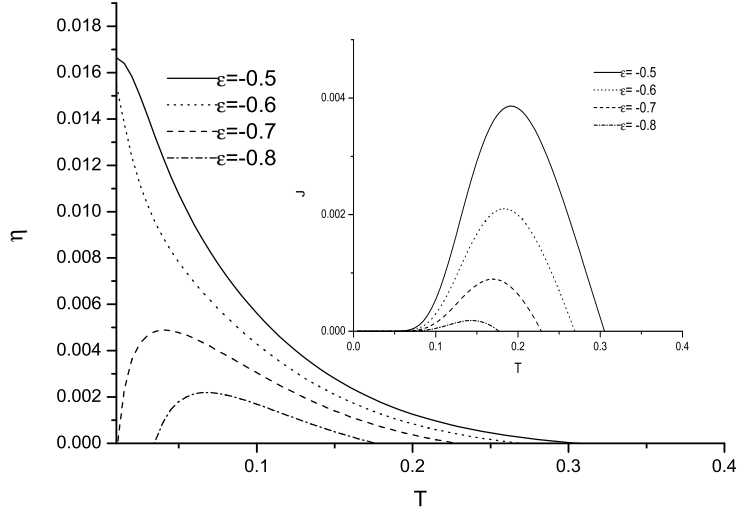}
  \caption{Efficiency $\eta$ as a function of temperature $T$ for different values of
  asymmetric parameters $\varepsilon=-0.5, -0.6, -0.7, -0.8$, $F=1.0$, $\lambda=1.0$
  ,$\Delta=0.7$, $Q=1.0$ and $L=0.01$. The inset shows the net current $J$ as function of $T$ for the same parameters.}\label{2}
\end{center}
\end{figure}
In Fig. 2 we plot the efficiency $\eta$ as a function of the
temperature for different values of $\varepsilon$
($\varepsilon<0$) with the parameter values, $F=1.0$,
$\lambda=1.0$, $\Delta=0.7$, $Q=1.0$ and $L=0.01$. From the figure
we can see that the efficiency is a decreasing function of
temperature for the cases of $\varepsilon=-0.5$ and
$\varepsilon=-0.6$, which shows that the presence of thermal
fluctuation dose not help efficient energy transformation by
ratchet. But for the cases of $\varepsilon=-0.7$ and
$\varepsilon=-0.8$ the efficiency can be optimized at finite
temperatures. In contradiction with the results of Ref (19) we
found a region where the efficiency attains a maximum at a finite
temperature. This shows that thermal fluctuations may facilitate
the energy conversion for asymmetric unbiased fluctuations. The
current is a peaked function of temperature for corresponding
parameters as shown in the inset. The highest temperature of the
ratchet decreases with the value of the asymmetric parameters
$\varepsilon$ of fluctuations and the lowest temperature of the
ratchet does not change with the $\varepsilon$, which indicates
that the asymmetric parameters are sensitive to the highest
working temperature of the ratchet. The peak shift to lower
temperature region with decreasing value of the asymmetric
parameters $\varepsilon$. Comparing Fig. 2 with the inset we can
see that the temperature corresponding to maximum current is not
the same as the temperature at which the efficiency is maximum,
which is consistent with the previous results
\cite{17}\cite{18}\cite{19}.

\indent From Eq. (16) we can know that the efficiency $\eta$
depends on the ratio $|\frac{j_{2}}{j_{1}}|$. If the function is
monotonically increasing, $\eta$ should be a monotonically
decreasing function of the temperature.
\begin{figure}[htbp]
  \begin{center}\includegraphics[width=11cm,height=8cm]{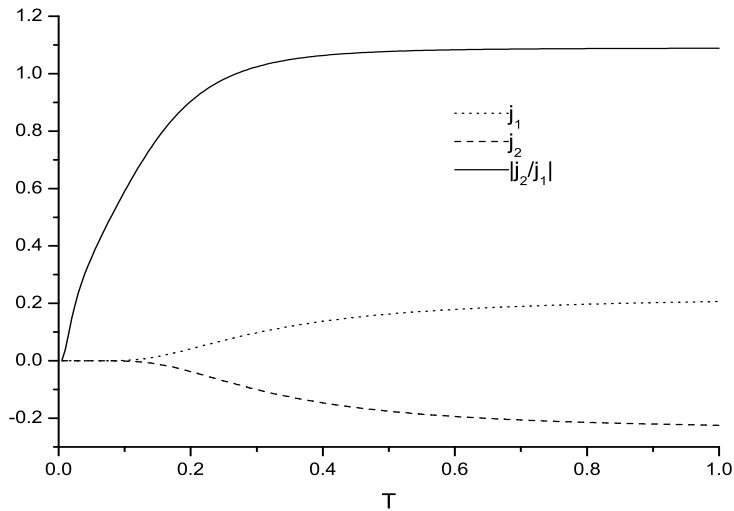}
  \caption{Plot of currents $j_{1}$, $j_{2}$ and $|\frac{j_{2}}{j_{1}}|$. The condition is the same as the case $\varepsilon=-0.5$ in Fig. 2}\label{3}
\end{center}
\end{figure}
\begin{figure}[htbp]
  \begin{center}\includegraphics[width=11cm,height=8cm]{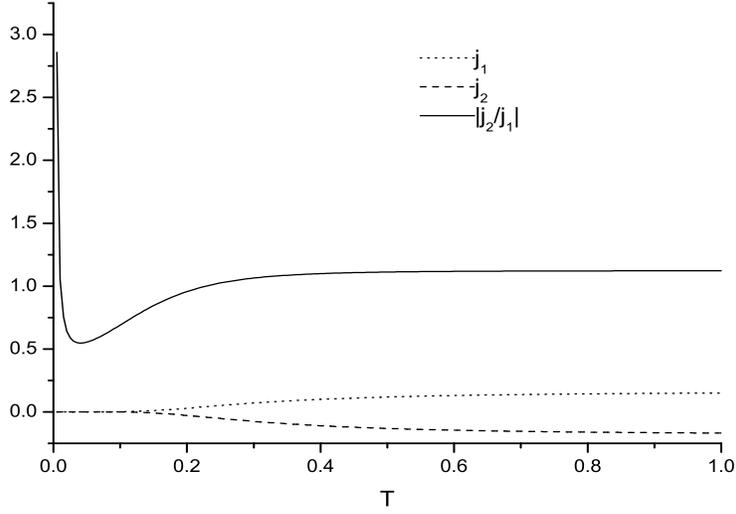}
  \caption{Plot of currents $j_{1}$, $j_{2}$ and $|\frac{j_{2}}{j_{1}}|$. The condition is the same as the case $\varepsilon=-0.7$ in Fig. 2}\label{4}
\end{center}
\end{figure}
In Fig. 3 we plot the fluxes obtained for the case of
$\varepsilon=-0.5$ (shown in Fig. 2). From Fig. 3 we can see that
the ratio $|\frac{j_{2}}{j_{1}}|$ is a monotonically increasing
function of temperature, which indicates that the efficiency
$\eta$ is decreasing function of temperature. However, for the
case of $\varepsilon=-0.7$ (see Fig. 4) the ratio
$|\frac{j_{2}}{j_{1}}|$ displays a clear minimum at the same value
of the temperature which corresponds to maximum of $\eta$ in Fig.
2.

\begin{figure}[htbp]
  \begin{center}\includegraphics[width=11cm,height=8cm]{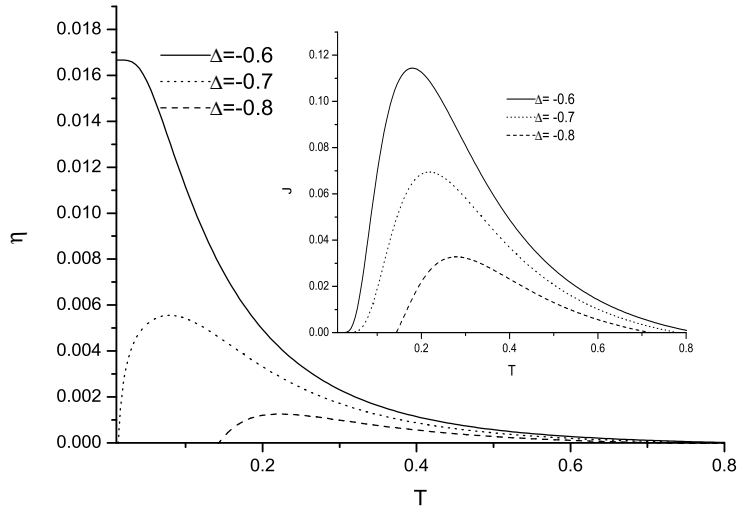}
  \caption{Efficiency $\eta$ as a function of temperature $T$ for different values of
  asymmetric parameters $\Delta=-0.6, -0.7, -0.8$, $F=1.0$, $\lambda=1.0$
  ,$\varepsilon=0.7$, $Q=1.0$ and $L=0.01$. The inset shows the net current $J$ as function of $T$ for the same parameters.}\label{2}
\end{center}
\end{figure}
In Fig. 5 we plot the efficiency $\eta$ as a function of the
temperature $T$ for different values of slope degree of potenial
$\Delta$ ($\Delta<0$) with the parameter values, $F=1.0$,
$\lambda=1.0$, $\varepsilon=0.7$, $Q=1.0$ and $L=0.01$. From the
figure, we can see that  with decreasing of $\Delta$ the
efficiency function of temperature becomes from a monotonically
decreasing function to a peaked function. This shows that the
thermal fluctuations actually facilitate the energy transformation
in some region.  The corresponding current is a peaked function of
temperature for the same parameters as shown in the inset. The
height of the peak decreases with the value of $\Delta$. The
lowest temperature of the ratchet changes with $\Delta$
drastically while the highest temperature of the ratchet does not
change with the $\Delta$ and the peak shift to higher temperature
region with decreasing value of $\Delta$, which is opposite to the
inset of the Fig. 2.
\begin{figure}[htbp]
  \begin{center}\includegraphics[width=11cm,height=8cm]{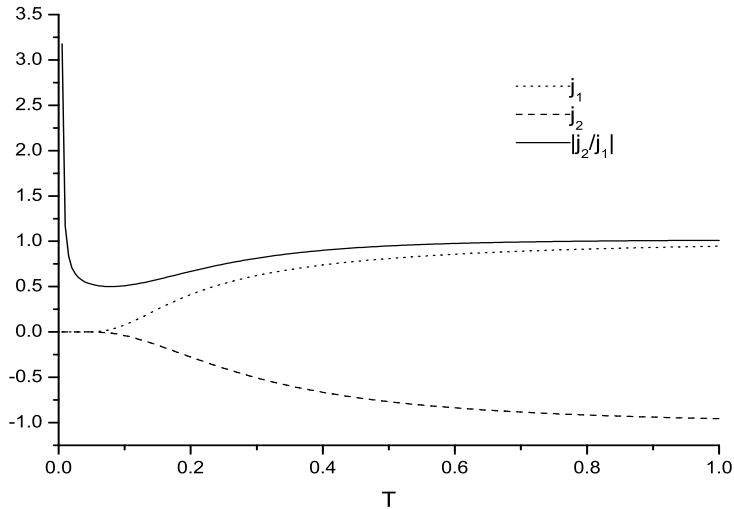}
  \caption{Plot of currents $j_{1}$, $j_{2}$ and $|\frac{j_{2}}{j_{1}}|$. The condition is the same as the case $\Delta=-0.7$ in Fig. 5}\label{2}
\end{center}
\end{figure}

\indent In Fig. 6, we plot the fluxes of the temperature for the
case of $\Delta=-0.7$ (shown in Fig. 5). From the Fig. 6 we can
see that the ratio $|\frac{j_{2}}{j_{1}}|$ displays a minimum at
the same value of the temperature , which  corresponds to maximum
of $\eta$ in Fig. 5\\
\section{4. Conclusion}
\indent In present paper, the transport of a Brownian particle
moving in spatially asymmetric potential in the presence of
asymmetric unbiased fluctuations is investigated.  In
contradiction with the previous findings, our results show that
the mutual interplay between asymmetry of fluctuation and
asymmetry of potential may cause an optimized efficiency of energy
conversion. This proves the claim made by Magnasco that there is a
region of the operating regime where the efficiency can be
optimized at finite temperatures. The temperature corresponding to
maximum current is not the same as the temperature at which the
efficiency is maximum. The asymmetry $\varepsilon$of fluctuation
is sensitive to the high temperature working region of the ratchet
while the asymmetry $\Delta$ of potential affects the low
temperature working region drastically.

\indent In this paper, the main features introduced by the
temporal asymmetry are the interplay of lower potential barriers
in positive direction relative to negative direction and the
corresponding shorter and longer times respectively the force is
felt. These type of competitive effects appear ubiquitously in
systems \cite{20} where there is an interplay between thermal
activation and dynamics.

\begin{acknowledgments}
The project supported by National Natural Science Foundation of
China (Grant No. of 10275099) and GuangDong Provincial Natural
Science Foundation (Grant No. of 021707 and 001182).
\end{acknowledgments}

\end{document}